\documentclass[amsmath,amssymb,aps,prd,nofootinbib,twocolumn,longbibliography]{revtex4-1}

\usepackage{graphicx}
\usepackage{hyperref}
\usepackage{xcolor}

\begin{document}

\title{Rotational-tidal phasing of the binary neutron star waveform}

\author{Philippe Landry}
\email{landryp@uchicago.edu}
\affiliation{Enrico Fermi Institute and Kavli Institute for Cosmological Physics, The University of Chicago, 5640 South Ellis Avenue, Chicago, Illinois, 60637, USA}

\date{\today}

\begin{abstract}
Tidal forces cause inspiralling binary neutron stars to deform, leaving a measurable imprint on the gravitational waves they emit. The induced stellar multipoles are an added source of gravitational radiation and modify the orbital dynamics, producing a slight acceleration of the coalescence which manifests as a phase shift in the waveform relative to point-particles. The dominant piece of this tidal phase comes from the mass quadrupoles, which contribute at fifth post-Newtonian order (5PN). Current quadrupoles and mass octupoles contribute at higher orders. For spinning neutron stars, additional multipole moments are induced by nonlinear couplings between spin and tides. We calculate these rotational-tidal deformations assuming the stars are rotating slowly and the tides are weak and quasi-stationary. The stellar multipole moments are read off from an asymptotically flat metric that encodes the difference between their tidal response and a black hole's. The multipoles are subsequently inserted into post-Newtonian formulas for the orbit and the gravitational radiation. We find that, at leading order, the rotational-tidal deformations make a 6.5PN contribution to the tidal phase. Their effect on the waveform is thus larger than that of the mass octupoles, and nearly as large as that of the current quadrupoles, in systems with non-negligible spin.
\end{abstract}

\maketitle

%%%%%%%%%%%%%%%%%%%%%%%%%%%%%%%%%%%%%%%%%%
\section{Introduction}\label{sec:intro}
%%%%%%%%%%%%%%%%%%%%%%%%%%%%%%%%%%%%%%%%%%

The LIGO-Virgo Collaboration's \cite{LIGO,VIRGO} detection of the gravitational waves from a binary neutron star merger, GW170817 \cite{GW170817}, recently established the first observational constraints on neutron-star tidal deformability. The tidal deformability $\lambda$ is an intrinsic property of a neutron star that is strongly correlated with the stiffness of its equation of state: for a fixed mass, a stiffer equation of state yields a larger value of $\lambda$. For the weak and slowly varying tides that occur during binary inspiral, $\lambda$ measures the size of the mass quadrupole

\begin{equation} \label{tiddef}
\mathcal{I}_{ab} = -\lambda \, \mathcal{E}_{ab}
\end{equation}
raised on one neutron star by the quadrupole moment $\mathcal{E}_{ab}$ of its companion's tidal field. (In Newtonian theory, the tidal quadrupole moment is related to the companion's gravitational potential, and hence its mass distribution, via $\mathcal{E}_{ab} = \partial_a\partial_b U^{\text{tid}}$, where $a, b$ are spatial indices. In general relativity, it stems from the electric part of the external spacetime's Weyl tensor.) Since the tidal deformability depends on the neutron star's internal structure, a measurement of $\lambda$ reveals information about the ultra-dense material at its core. Constraining the equation of state of matter in this high-density regime with gravitational-wave observations is an area of robust and ongoing activity \cite{Lackey,Annala,Chatziioannou,Radice,Agathos,DelPozzo,Markakis,HindererLackey,Maselli,Raithel,Harry,Most}.

Finite-size effects such as tides are inferred from a detected binary neutron star waveform by matching the data to a waveform model and performing Bayesian parameter estimation \cite{LALinf}. Waveform models (e.g.~\cite{Sathya,Buonanno,Arun}) for the inspiral stage of the coalescence, when the adiabatic tides captured by Eq.~\eqref{tiddef} are important, have been developed from post-Newtonian (PN) calculations of the binary's orbital and radiative dynamics (see Ref.~\cite{Blanchet} for a review). Calculations incorporating linear tides \cite{Flanagan,HindererLackey} have demonstrated that $\lambda$ contributes to the phasing of the waveform at fifth post-Newtonian order (5PN)---i.e.~with a factor of $(v/c)^{10}$ in a slow-motion expansion in powers of the orbital velocity---relative to the leading point-particle terms. This tidal phase correction stems from the slight acceleration of the coalescence engendered by two effects of the induced stellar quadrupoles: an enhancement of the relative acceleration of the neutron stars, as their deformations concentrate more mass on the axis that connects them; and an increase in the system's gravitational-wave energy flux, as the tidal bulges source additional radiation through the quadrupole formula. Because each neutron star contributes its own tidal deformability $\lambda^{(i)}, i=1,2$, the 5PN tidal phase actually involves the weighted average \cite{Flanagan,HindererLackey}

\begin{equation}
 \tilde{\Lambda} = \frac{16}{13}c^{10} \frac{ (1+12q)\lambda^{(1)}+(1+12/q)\lambda^{(2)} }{M^5} ,
 \end{equation} 
where $q:=m_2/m_1$ is the mass ratio and $M := m_1 + m_2$ is the total mass of the binary.\footnote{We label the neutron stars such that $m_2 \leq m_1$.} Constraints from GW170817 place an upper bound of $\tilde{\Lambda} \leq 800$ on this effective tidal deformability at 90\% confidence \cite{GW170817}. The deformabilities $\lambda^{(1)}$ and $\lambda^{(2)}$ also appear in a different combination \cite{Favata,Wade},

\begin{align}
\delta \tilde{\Lambda} =& \; c^{10} \frac{(1319-7996q-11005q^2)\lambda^{(1)}}{1319M^6} \nonumber \\ &-c^{10} \frac{(1319-7996/q-11005q^2)\lambda^{(2)}}{1319M^6} ,
\end{align}
at next-to-leading order (6PN). The parameter $\delta\tilde{\Lambda}$ is more difficult to measure \cite{Wade} and was not reported in Ref.~\cite{GW170817}.

Although the deformations generated by $\mathcal{E}_{ab}$ make the dominant contribution to the tidal phase, they are not the only tidal deformations that arise during inspiral. Higher-multipole tidal moments sourced by the companion's mass distribution also imprint on the waveform. Generalizing Eq.~\eqref{tiddef}, they raise mass $\ell$-poles \cite{Damour_Nagar,Binnington}

\begin{equation} \label{tiddef}
\mathcal{I}_L = -\lambda_{\ell} \, \mathcal{E}_L
\end{equation}
on the neutron star, where $L := a_1 a_2 ... a_{\ell}$ is an $\ell$-fold spatial multi-index. ($\mathcal{E}_L = \partial_L U^{\text{tid}}$ in Newtonian theory.) Ref.~\cite{Yagi} has shown that the tidal deformabilities $\lambda_{\ell}$ contribute to the waveform phase at $(2\ell+1)$PN.

The tidal moments $\mathcal{E}_L$ are called \emph{gravitoelectric}, in contradistinction to \emph{gravitomagnetic} tidal moments $\mathcal{B}_L$ that induce current, rather than mass, multipoles. The gravitomagnetic tidal moments are sourced by the companion's momentum distribution (i.e.~its orbital motion), and unlike $\mathcal{E}_L$ they have no Newtonian analogue, being a purely relativistic phenomenon. Gravitomagnetic tides in neutron-star binaries have been investigated by several authors \cite{Favata,Landry_irrot,Yagi,Damour_Nagar,Binnington}. In particular, Refs.~\cite{Damour_Nagar,Binnington} introduced gravitomagnetic counterparts $\sigma_{\ell}$ to the gravitoelectric tidal deformabilities $\lambda_{\ell}$, defined here by the linear relation

\begin{equation} \label{magdef}
\mathcal{S}_L = - \sigma_{\ell} \mathcal{B}_L
\end{equation}
between the induced current multipoles $\mathcal{S}_L$ and the tidal moments. Ref.~\cite{Yagi} calculated the effect of $\sigma$ on the waveform phase,\footnote{We suppress the subscript $\ell$ on $\lambda_{\ell}$, $\sigma_{\ell}$ when referring to quadrupole deformabilities, such that $\lambda := \lambda_2$ and $\sigma := \sigma_2$. We indicate the label $\ell$  explicitly on higher-multipole deformabilities.} finding that it enters at 6PN---the same order as $\delta\tilde{\Lambda}$---in the combination

\begin{align}
\tilde{\Sigma} = \frac{10}{21} c^{10} \frac{(1-1037q)\sigma^{(1)} + (1-1037/q)\sigma^{(2)}}{M^5} .
\end{align}
It was originally thought that $\tilde{\Sigma}$ vanishes in equal-mass binaries, but this claim has recently been overturned \cite{Yagi,Yagi_err}.

Because general relativity is a nonlinear theory, in principle the tidal moments $\mathcal{E}_L$ and $\mathcal{B}_L$ couple to one another to produce nonlinear tides. Similar couplings between the tidal moments and the neutron star's spin also occur. In this paper, we focus exclusively on the spin-tide couplings, which can be significant even when the stars are rotating slowly \cite{Pani_ext,Landry_ext,GagnonBischoff}. The composition of the dipole spin vector $S^a$ and the quadrupole ($\mathcal{E}_{ab},$ $\mathcal{B}_{ab}$) or octupole ($\mathcal{E}_{abc},$ $\mathcal{B}_{abc}$) tidal moments produces corrections

\begin{subequations} \label{rottiddef}
\begin{align}
\delta \mathcal{I}_{ab} &= -\hat{\lambda}_2 \chi^c \mathcal{B}_{abc} , \\
\delta \mathcal{S}_{ab} &= \hat{\sigma}_2 \chi^c \mathcal{E}_{abc} , \\
\delta \mathcal{I}_{abc} &= -\hat{\lambda}_3 \chi_{\langle c} \mathcal{B}_{ab \rangle} , \\
\delta \mathcal{S}_{abc} &= \hat{\sigma}_3 \chi_{\langle c} \mathcal{E}_{ab \rangle}
\end{align}
\end{subequations}
to the tidally induced multipoles\footnote{In this study, we ignore the $O(\chi^2)$ spin quadrupole because the star's rotation is assumed to be slow. The $O(\chi)$ rotational-tidal hexadecapole ($\ell = 4$) moments $\delta \mathcal{I}_{abcd} = -\hat{\lambda}_4 \chi_{\langle d} \mathcal{B}_{abc \rangle}$, $\delta \mathcal{S}_{abcd} = \hat{\sigma}_4 \chi_{\langle d} \mathcal{B}_{abc \rangle}$ that arise from the spin-octupole couplings are also omitted, as we choose to truncate at $\ell=3$. The rotational-tidal dipole moments induced by spin-quadrupole couplings do not appear in Eq.~\eqref{rottiddef}, since they represent an overall acceleration of the neutron star.} at first order in the dimensionless spin $\chi^a := cS^a/GM^2$ \cite{Pani,Landry_ext,GagnonBischoff}. (Angular brackets denote symmetrization and trace-removal.) These \emph{rotational-tidal} deformations have been the subject of a number of recent studies \cite{Pani_ext,Landry_ext,Pani,Landry_dynresp,Landry_int,GagnonBischoff}. Observe that because $\chi^a$ is a pseudovector, spin-coupled gravitoelectric tides give rise to current multipoles, and spin-coupled gravitomagnetic tides give rise to mass multipoles. We have introduced rotational-tidal deformabilities $\hat{\lambda}_{\ell}$, $\hat{\sigma}_{\ell}$ to measure the amplitude of the multipoles $\delta \mathcal{I}_L$ and $\delta\mathcal{S}_L$, respectively. These quantities are closely related to the \emph{rotational-tidal Love numbers} discussed in the aforementioned works.

The corrections \eqref{rottiddef} are expected to manifest themselves in the waveform's phasing \cite{Pani,GagnonBischoff}, but no analysis of their impact has been performed to date. While Ref.~\cite{GagnonBischoff} presented the metric outside a deformed, slowly rotating neutron star in a weak-field expansion that permitted the relative sizes of different rotational-deformations to be compared, their phase contributions were not addressed. Similarly, Ref.~\cite{Pani} calculated a subset of the tidally induced Geroch-Hansen multipole moments explicitly using Ryan's method  \cite{Ryan1,Ryan2}, but did not consider the gravitational waveform. In this paper, we complete the pioneering work of Refs.~\cite{Pani,GagnonBischoff} by computing the leading-order waveform phasing produced by the rotational-tidal deformations.

Broadly speaking, the strategy is to take the metric of a deformed, rotating neutron star, remove the terms associated with the external tidal field (retaining those associated with the tidal response) and read off the multipole moments. The multipoles are then incorporated into PN formulas for the orbit and the gravitational-wave energy flux; the modifications relative to the inspiral of two point-particles produce the tidal phase. We now flesh out this approach in greater detail.

We calculate all the $\ell \leq 3$ multipole moments of a slowly rotating neutron star deformed by quadrupole and octupole tides, working to first order in deformations and in the dimensionless spin $\chi := |\chi^a|$. (In practice, this means relating the rotational-tidal deformabilities of Eq.~\eqref{rottiddef} to the Love numbers known to appear in the metric \cite{Pani_ext,Landry_ext}, which can be computed with published methods \cite{Pani,Landry_int,GagnonBischoff}.) The multipoles are obtained using a different prescription than Ref.~\cite{Pani_ext}'s: inspired by Ref.~\cite{Gralla}, we rely on an asymptotically flat difference metric that encodes how the neutron star's tidal response differs from that of black hole. Because black holes have zero tidal response \cite{Binnington,Landry_ext}, this difference spacetime naturally retains the radially decaying tidal response while removing the radially growing external tidal field. The asymptotic flatness of the difference metric permits the quadrupoles and octupoles to be read off from the multipole structure of Thorne's generic metric for stationary, asymptotically flat vacuum spacetimes in asymptotically Cartesian mass-centred (ACMC) coordinates \cite{Thorne}. While our expressions for the multipole moments should necessarily be the same as the subset calculated in Ref.~\cite{Pani}, the value of our approach is that it avoids the ambiguities identified in Ref.~\cite{Gralla} and Sec.~III F of Ref.~\cite{Pani_ext}. Moreover, Ref.~\cite{Gralla} argues that it is precisely such a \emph{difference} in tidal responses that is physically meaningful.

With the multipoles in hand, the tidal corrections to the frequency-domain gravitational-wave phase $\Psi$ are calculated in the stationary-phase approximation \cite{Cutler} for an aligned-spin circular binary. The computation parallels Ref.~\cite{Flanagan}'s seminal calculation of the 5PN $\tilde{\Lambda}$ term (see also Refs.~\cite{Kochanek,HindererLackey,Yagi}), with some modifications to account for the spin-tide couplings. In particular, we must account for current octupole radiation and include a 1PN correction to the binary's mass quadrupole \cite{Yagi_err}. The net effect of these modifications is to slightly augment both the relative acceleration and the radiative dissipation in the binary. The rotational-tidal deformations consequently produce a modest speed-up of the coalescence in excess of the known effect from linear tides, thereby increasing the tidal phase by a small amount.

The final result of the calculation is the following expression for the gravitational-wave phase $\Psi$, expanded in terms of the PN parameter $x:=(v/c)^2$ and neglecting next-to-leading-order terms in each tidal deformability:

\begin{align} \label{psifinal}
\Psi = \frac{3M}{128 \mu}x^{-2.5} \bigg[ 1 &- \frac{39}{2}\tilde{\Lambda} \, x^5 + \tilde{\Sigma} \, x^6 - \tilde{X} \, x^{6.5} \nonumber \\ &- \tilde{\Lambda}_3 \, x^7 + \tilde{\Sigma}_3 \, x^8 \bigg] ,
\end{align}
with the rotational-tidal phase contribution given by

\begin{widetext}
\begin{align} \label{X}
\tilde{X} =& \frac{1}{21M^6} c^{12} \bigg\{ \chi^{(1)} \left[ 36(35+614q)\hat{\lambda}_2^{(1)} - (7-4751q)\hat{\sigma}_2^{(1)} - 2316 q \hat{\lambda}_3^{(1)} - 3474 q \hat{\sigma}_3^{(1)} \right] \nonumber \\ &+ \chi^{(2)} \left[ 36(35+614/q)\hat{\lambda}_2^{(2)} - (7-4751/q)\hat{\sigma}_2^{(2)} - 2316 \hat{\lambda}_3^{(2)}/q - 3474 \hat{\sigma}_3^{(2)}/q \right] \bigg\} ,
\end{align}
\end{widetext}
and the octupole phase terms given by

\begin{align}
\tilde{\Lambda}_3 &= \frac{4000}{9 M^7}c^{14}(q \lambda_3^{(1)}+ \lambda_3^{(2)}/q) , \label{lambda3} \\
\tilde{\Sigma}_3 &=  \frac{29925}{11 M^7}c^{14} (q \sigma_3^{(1)}+\sigma_3^{(2)}/q) . \label{sigma3}
\end{align}
Here, $\mu := m_1 m_2/M$ is the binary's reduced mass, and superscripts (1), (2) label the individual neutron stars. We emphasize that this expression omits all PN point-particle phase terms (which are currently known up to 3.5PN \cite{Blanchet}), as well as next-to-leading-order corrections (such as $\delta \tilde{\Lambda}$) to each tidal term. The purpose of this work is to determine the most important rotational-tidal phasing terms, rather than attempt a consistent derivation of all tidal terms up to e.g.~6.5PN.\footnote{We remark that the calculation presented here fails to recover the exact phasing coefficient $\tilde{\Sigma}$ as defined above. The disagreement with Ref.~\cite{Yagi_err} is discussed in Sec.~\ref{sec:disc}.}

The PN scaling of Eq.~\eqref{psifinal} indicates that the rotational-tidal phasing of the waveform enters at 6.5PN. It is therefore suppressed by a factor of $\sim\chi (v/c)^3$ relative to the dominant 5PN tidal term proportional to $\tilde{\Lambda}$, but it is only a factor of $\sim\chi (v/c)$ smaller than the 6PN terms proportional to $\tilde{\Sigma}$ and $\delta{\tilde{\Lambda}}$. Furthermore, the rotational-tidal phase contribution is in general larger than the corrections due to the tidal octupole moments, which enter at 7PN and 8PN for $\tilde{\Lambda}_3$ and $\tilde{\Sigma}_3$, respectively. Since $\tilde{X}$ scales linearly with the dimensionless spins of the neutron stars, the rotational-tidal phase may be important in systems with significant spin.\footnote{The slow rotation approximation is valid even for the fastest-rotating neutron stars observed in a binary ($\chi \sim 0.05$). For very rapidly rotating compact objects, like millisecond pulsars ($\chi \sim 0.5$), one expects $O(\chi^2)$ corrections ignored in this work to become important.} Remarkably, all four rotational-tidal couplings identified in Eq.~\eqref{rottiddef} turn out to contribute to $\Psi$ at the same PN order.

The phasing given in Eq.~\eqref{psifinal} is worked out in detail in the remainder of the paper, which is organized as follows. Sec.~\ref{sec:metric} presents the metric outside a tidally deformed, slowly rotating neutron star. The neutron star's multipole moments are calculated in Sec.~\ref{sec:multipole}. Sec.~\ref{sec:tidphase} is devoted to computing the phase correction produced by each of these multipoles. Finally, the implications of the results summarized in this introduction are discussed in Sec.~\ref{sec:disc}.

Throughout the paper, spatial indices are denoted with lowercase Latin characters $a,b,c,... \,$, while spacetime indices are denoted with lowercase Greek characters $\alpha,\beta,\gamma,... \,$. Uppercase Latin indices $A,B,C,...$ refer to angular variables $\theta^A :=(\theta,\phi)$ on the two-sphere. The spatial indices are raised and lowered with the Euclidean metric $\delta_{ab}$ and its inverse. Geometrized $G=c=1$ units are employed in Sec.~\ref{sec:tidphase}.

%%%%%%%%%%%%%%%%%%%%%%%%%%
\section{Spacetime of a tidally deformed, slowly rotating neutron star} \label{sec:metric}
%%%%%%%%%%%%%%%%%%%%%%%%%%

The metric outside a tidally perturbed, rotating neutron star of mass $m$, radius $R$ and dimensionless spin vector $\chi^a$ was derived in Refs.~\cite{Landry_ext} and \cite{Pani}, assuming that both the deformations from sphericity and the dimensionless spin $\chi$ are small. Ref.~\cite{Pani} considered the influence of the quadrupole ($\ell = 2$) and octupole ($\ell = 3$) moments of the external tidal field, which was idealized as stationary---as per the usual adiabatic approximation for equilibrium tides (see e.g.~Sec.~I of Ref.~\cite{Poisson_bh})---and axisymmetric, but otherwise generic. The restriction to axial symmetry was dropped by Ref.~\cite{Landry_ext}, which nonetheless only considered the tidal quadrupole moments. Ref.~\cite{Landry_ext} worked to first order in $\chi$, while Ref.~\cite{Pani} worked to second order.

Here, we adopt the notation and approximations of Ref.~\cite{Landry_ext}, but add in the octupole-generated terms derived in Ref.~\cite{Pani}. Accordingly, the metric we employ describes to first order in spin and in deformations the spacetime of a neutron star subject to quadrupole and octupole tidal fields. In particular, we retain bilinear terms that are manifestations of rotational-tidal couplings at lowest perturbative order.

The metric is expressed in Cartesian coordinates $x^a$ in the Regge-Wheeler gauge \cite{Landry_ext}.
The Euclidean distance from the neutron star's centre of mass is $r$, and $n^a := x^a/r$ is the radial unit vector; in spherical coordinates, $n^a = [\sin{\theta}\cos{\phi},\sin{\theta}\sin{\phi},\cos{\theta}]$. Following Ref.~\cite{GagnonBischoff}, we expand the metric in powers of $Gm/c^2 r$. Its time-time ($tt$) and time-angle ($tA$) components take the form

\begin{widetext}
\begin{subequations} \label{nsmetric}
\begin{align}
g_{tt} =& - 1 + \frac{2Gm}{c^2 r} - \left[ 1 + ... + 2k_2^{\text{el}} \left( \frac{R}{r} \right)^5 \left( 1 + ...\right) \right] \frac{\mathcal{E}_{ab} x^a x^b}{c^2} -\frac{1}{3} \left[ 1 + ... + 2k_3^{\text{el}} \left( \frac{R}{r} \right)^7 \left( 1 + ...\right) \right] \frac{\mathcal{E}_{abc} x^a x^b x^c}{c^2} \nonumber \\ &+\frac{2Gm}{c^2}(1+...)\frac{\chi^b\mathcal{B}_{ab}x^a}{c^3} - \frac{24 Gm}{35 c^2} \left[ \frac{G m}{c^2 r} + ... + 2 \mathfrak{k}^{\text{q}} \left( \frac{R}{r} \right)^5 \left( 1 + ... \right)  \right] \frac{\chi^{c} \mathcal{B}_{abc} x^a x^b}{c^2} \nonumber \\ &- \frac{2Gm}{c^2 r^2} \left[ \frac{Gm}{c^2 r} + ... + 2 \mathfrak{k}^{\text{o}} \left( \frac{R}{r} \right)^5 \left( 1 + ... \right)  \right] \frac{\chi_{\langle a} \mathcal{B}_{bc \rangle} x^a x^b x^c}{c^3} , \\
g_{tA} =& \frac{2G^2 m^2}{c^4 r^3} \epsilon_{abc} \chi^c x^b x^a_A + \frac{2}{3} \left[ 1 + ... - 6 \left(\frac{Gm}{c^2 r} \right) k_2^{\text{mag}} \left(\frac{R}{r} \right)^4 \left( 1 + .. \right) \right] \frac{\epsilon_{acd} x^c \mathcal{B}^d_{\;\,b} x^b x^a_A}{c^3} \nonumber \\ &+ \frac{1}{3} \left[ 1 + ... - \frac{16}{3} \left(\frac{Gm}{c^2 r} \right) k_3^{\text{mag}} \left(\frac{R}{r} \right)^6 \left( 1 + .. \right) \right] \frac{\epsilon_{acd} x^c \mathcal{B}^d_{\;\,be} x^b x^e x^a_A}{c^4} \nonumber \\ &-\frac{2G^2m^2}{c^4 r}(1+...)\frac{\epsilon_{abc}x^b\mathcal{E}^c_{\;\;d}\chi^dx^a_A}{c^2}- \frac{10 G m}{3 c^2} \left[ \frac{G m}{c^2 r} + ... + \frac{3}{5}\mathfrak{f}^{\text{q}} \left( \frac{R}{r} \right)^5 \left( 1 + ... \right) \right] \frac{\epsilon_{acd} x^c \mathcal{E}^d_{\;\;be} \chi^{e} x^b x^a_A}{c^3} \nonumber \\ &- \frac{10Gm}{3c^2 r^2} \left[ \frac{G^2m^2}{c^4 r^2} + ... + \frac{6}{5}\mathfrak{f}^{\text{o}} \left( \frac{R}{r} \right)^5 \left( 1 + ... \right) \right] \frac{\epsilon_{ac}^{\;\;\,\,d} \mathcal{E}_{\langle db} \chi_{e\rangle} x^c x^b x^e x^a_A}{c^2} ,
\end{align}
\end{subequations}
\end{widetext}
where $\epsilon_{abc}$ is the Levi-Civita symbol and ellipses denote relativistic corrections of order $Gm/c^2 r$ and higher. The perturbative structure of the metric is clearly recognizable: the first two terms in $g_{tt}$ represent the Schwarzschild background describing an isolated, non-rotating neutron star; the first term in $g_{tA}$ accounts for the star's slow rotation; the terms proportional to the tidal moments $\mathcal{E}_L$ or $\mathcal{B}_L$ for $\ell = 2, 3$ represent quadrupole and octupole tidal deformations; and the terms proportional to a combination of $\chi^a$ and the tidal moments encode dipole, quadrupole and octupole rotational-tidal deformations due to spin-tide couplings. Each set of terms enclosed in square brackets consists (when the outer factors of $x^a=r n^a$ are taken into account) of a radially growing piece---the external tidal field---and a radially decaying piece---the tidal response---whose amplitude is measured by a \emph{Love number}. The (gravitoelectric) tidal Love numbers $k_{\ell}^{\text{el}}$ characterize the response to the applied gravitoelectric $\ell$-poles $\mathcal{E}_L$ and are closely related to $\lambda_{\ell}$; the gravitomagnetic Love numbers $k_{\ell}^{\text{mag}}$ do likewise for the gravitomagnetic $\ell$-poles $\mathcal{B}_L$ and are related to $\sigma_{\ell}$; and the rotational-tidal Love numbers $\mathfrak{f}^{\text{q}}, \mathfrak{f}^{\text{o}}, \mathfrak{k}^{\text{q}}$ and $\mathfrak{k}^{\text{o}}$ characterize the responses to the bilinear moments $\epsilon_{acd} \mathcal{E}^d_{\;\; bc} \mathcal{\chi}^c, \epsilon_{ac}^{\;\;\; d} \mathcal{E}_{\langle db} \chi_{c\rangle}, \epsilon_{acd} \mathcal{B}^d_{\;\; bc} \mathcal{\chi}^c$ and $\epsilon_{ac}^{\;\;\; d} \mathcal{B}_{\langle db} \chi_{c\rangle}$, respectively. The rotational-tidal Love numbers $\mathfrak{f}^{(\ell)}$ and $\mathfrak{k}^{(\ell)}$ are respectively related to $\hat{\sigma}_{\ell}$ and $\hat{\lambda}_{\ell}$. There are no Love numbers associated with the dipole terms in Eq.~\eqref{nsmetric}, as these represent an overall acceleration of the neutron star due to spin forces \cite{GagnonBischoff}. The Love numbers appear as unknown integration constants in the metric, and to calculate them it is necessary to solve for the internal structure of the perturbed neutron star (see Refs.~\cite{Pani,Landry_int,GagnonBischoff}). To establish the Love numbers' precise relations to the tidal deformabilities, one must compute the multipole moments \eqref{tiddef}, \eqref{magdef} and \eqref{rottiddef} for Eq.~\eqref{nsmetric}.

The tidal moments are related to the components of the asymptotic Weyl tensor sourced by external mass and momentum distributions \cite{Zhang}. For a circular binary, the tidal field is generated by the companion body (labeled (2)), and its $\ell = 2, 3$ multipole moments take the form \cite{Poisson_bh}

\begin{align}
\mathcal{E}_{L} &= - (-1)^{\ell} (2\ell - 1)!! Gm_2\frac{n_{\langle L \rangle}}{r^{\ell+1}} , \\
\mathcal{B}_{L} &= - (-1)^{\ell} (2\ell - 1)!! \ell! Gm_2 \epsilon_{bc\langle a_{\ell}} \frac{n_{L-1 \rangle b}}{r^{\ell+1}} v^{c}
\end{align}
at leading PN order, namely 0PN for $\mathcal{E}_L$ and 1PN for $\mathcal{B}_L$. Here, the combination $n_L$ stands for the product $n_{a_1} n_{a_2} ... n_{a_{\ell}}$. In this specialized setting, the vector $n_a$ points from neutron star 1 to neutron star 2 as it orbits at velocity $v$ and angular frequency $\Omega$, such that $\theta=\pi/2$ and $\phi=\Omega t$. The tidal moments are symmetric and trace-free (STF); for the properties of such tensors, and for identities involving STF combinations of $n_a$, the reader is referred to Appendix A of Ref.~\cite{Racine}.

Refs.~\cite{Binnington, Landry_ext} demonstrated that the tidal, gravitomagnetic and rotational-tidal Love numbers vanish for black holes, in contrast to material bodies. Consequently, the metric of a tidally deformed, slowly rotating black hole \cite{Poisson_bh} can be recovered from Eq.~\eqref{nsmetric} by setting the Love numbers to zero.

%%%%%%%%%%%%%%%%%%%%%%%%%%
\section{Multipole moments} \label{sec:multipole}
%%%%%%%%%%%%%%%%%%%%%%%%%%

The general-relativistic multipole moments for a non-spherical compact object can be determined from the far-field behaviour of the spacetime metric. Ref.~\cite{Thorne} derived the general form of the metric of a stationary, asymptotically flat vacuum spacetime in an ACMC-coordinate based multipole expansion. Provided one's spacetime fits the bill, one can transform to ACMC coordinates and simply read off the multipole moments by comparing with Eq.~(11.1) of Ref.~\cite{Thorne}.

We wish to implement this procedure to calculate the multipole moments for the deformed, spinning neutron star of Eq.~\eqref{nsmetric}. However, although it is a stationary vacuum solution, this metric is manifestly non-flat at spatial infinity because of the radially growing terms associated with the external tidal field. While it would be impossible to disentangle the applied field from the tidal response in the full nonlinear theory of gravity, the tidal deformations are defined in a perturbation theory linearized about a Schwarzschild background, and consequently there exists a superposition principle for the perturbations. Indeed, if $g_{\alpha\beta}^{\text{NS}} := g_{\alpha\beta}^0 + p_{\alpha\beta}^{\text{NS}}$ and $g_{\alpha\beta}^{\text{BH}} := g_{\alpha\beta}^0 + p_{\alpha\beta}^{\text{BH}}$ are each solutions to the Einstein field equations linearized about the Schwarzschild solution $g_{\alpha\beta}^0$, it follows that the difference metric

\begin{equation}
h_{\alpha\beta} = g_{\alpha\beta}^{0} + \delta p_{\alpha\beta} , \qquad \delta p_{\alpha\beta} := p_{\alpha\beta}^{\text{NS}} - p_{\alpha\beta}^{\text{BH}}
\end{equation}
---obtained by identifying the background metrics and taking the perturbations to live on the same manifold---is also a solution \cite{Gralla}. Because $g_{\alpha\beta}^{\text{BH}}$ contains only growing applied tidal field terms (black holes have vanishing Love numbers), $h_{\alpha\beta}$ contains only decaying tidal response terms. The difference metric is thus precisely the quantity of interest: the asymptotically flat metric representing the tidal response, without the applied tidal field. Its multipole moments describe how the structure of a tidally deformed neutron star differs from the structure of an (intrinsically zero-response) black hole deformed by the same tidal field.\footnote{In our case, we take $p_{\alpha\beta}^{\text{NS}}$ and $p_{\alpha\beta}^{\text{BH}}$ to be perturbations of the same Schwarzschild background that include the $O(\chi)$ rotational term, the linear tidal terms and the $O(\chi)$ rotational-tidal terms. The scheme would not work as neatly at $O(\chi^2)$, because the spin quadrupole terms that appear at this order are different for neutron stars and black holes.}

In principle, one could compute $h_{\alpha\beta}$ in any coordinate system, provided the perturbations $p_{\alpha\beta}^{\text{NS}}$ and $p_{\alpha\beta}^{\text{BH}}$ are expressed in the same gauge, and then transform the result to ACMC coordinates. In our case, the coordinates of Eq.~\eqref{nsmetric} are already ACMC after removal of the growing terms, and the black hole metric is obtained by trivially setting the Love numbers to zero, so we are spared the coordinate transformation. The relevant components of the difference metric $h_{\alpha\beta}$ are thus

\begin{subequations} \label{diffmetric}
\begin{align}
h_{tt} =& -1 + \frac{2Gm}{c^2 r} - 2k_2^{\text{el}} \left( \frac{R}{r} \right)^5 \frac{\mathcal{E}_{ab} x^a x^b}{c^2} \nonumber \\ &- \frac{2}{3}k_3^{\text{el}} \left( \frac{R}{r} \right)^7  \frac{\mathcal{E}_{abc} x^a x^b x^c}{c^2} \nonumber \\ &- \frac{48G m}{35c^2} \mathfrak{k}^{\text{q}} \left(\frac{R}{r} \right)^5 \frac{\chi^{c} \mathcal{B}_{abc} x^a x^b}{c^2} \nonumber \\ &- \frac{4Gm}{c^2 r^2} \mathfrak{k}^{\text{o}} \left( \frac{R}{r} \right)^5  \frac{\chi_{\langle a} \mathcal{B}_{bc \rangle} x^a x^b x^c}{c^3} , \\
h_{tA} =& \frac{2G^2m^2}{c^4r^3}\epsilon_{abc}\chi^c x^b x^a_A \nonumber \\ &-4 \left(\frac{Gm}{c^2 r} \right) k_2^{\text{mag}} \left(\frac{R}{r} \right)^4 \frac{\epsilon_{acd} x^c \mathcal{B}^d_{\;\,b} x^b x^a_A}{c^3} \nonumber \\ &-\frac{16}{9} \left(\frac{Gm}{c^2 r} \right) k_3^{\text{mag}} \left(\frac{R}{r} \right)^6 \frac{\epsilon_{acd} x^c \mathcal{B}^d_{\;\,be} x^b x^e x^a_A}{c^4} \nonumber \\ &-\frac{2Gm}{c^2} \mathfrak{f}^{\text{q}} \left(\frac{R}{r} \right)^5 \frac{\epsilon_{acd} x^c \mathcal{E}^d_{\;\;be} \chi^{e} x^b x^a_A}{c^3} \nonumber \\ &- \frac{4Gm}{c^2 r^2} \mathfrak{f}^{\text{o}} \left( \frac{R}{r} \right)^5 \frac{\epsilon_{ac}^{\;\;\,\,d} \mathcal{E}_{\langle db} \chi_{e\rangle} x^c x^b x^e x^a_A}{c^2} .
\end{align}
\end{subequations}
We have dispensed with the ellipses denoting higher-order relativistic corrections, as it is understood that this metric represents only the leading-order PN terms associated with the various tidal deformations. Since $h_{\alpha\beta}$ satisfies all the criteria of Ref.~\cite{Thorne} by construction, we may compare it directly to the metric given in Eq.~(11.1) of that reference, which we rewrite as

\begin{subequations} \label{Thornemetric}
\begin{align}
h_{tt} =& -1 + \frac{2G \mathcal{I}}{c^2 r} + \frac{[\text{0-pole}]}{r^2} + \sum_{\ell = 2}^{\infty} \frac{1}{r^{\ell+1}} \bigg[\frac{2(2\ell-1)!!}{\ell!}\frac{G}{c^2} \mathcal{I}_L n^L \nonumber \\ &+ [(\ell-1)\text{-pole}] + ... + [\text{0-pole}] \bigg] \\
h_{tA} =& \sum_{\ell = 1}^{\infty} \frac{1}{r^{\ell+1}} \bigg[-\frac{4\ell(2\ell-1)!!}{(\ell+1)!} \frac{G}{c^3} \epsilon_{abc} \mathcal{S}_{b(L-1)}n^{c(L-1)}x^a_A \nonumber \\ &+ [(\ell-1)\text{-pole}] + ... + [\text{0-pole}] \bigg]  .
\end{align}
\end{subequations}
(Here we have transformed Ref.~\cite{Thorne}'s time-space component into a time-angle component via $h_{tA} = h_{ta} \partial_A x^a$, and restored factors of $G$ and $c$.) Because we have suppressed the relativistic corrections in Eq.~\eqref{nsmetric}, only the leading-order terms proportional to $\mathcal{I}_L$ or $\mathcal{S}_L$ in Eq.~\eqref{Thornemetric} have counterparts in Eq.~\eqref{diffmetric}.

The multipole moments determined by comparing Eqs.~\eqref{diffmetric} and \eqref{Thornemetric} are as follows. The (mass) monopole and (current) dipole moments

\begin{equation}
\mathcal{I} = m , \qquad \mathcal{S}_a = G m^2 \chi_a/c
\end{equation}
are interpreted in the usual way as the mass and spin angular momentum $S_a$ of the unperturbed neutron star. The stellar mass and current multipoles induced by the tides are

\begin{equation} \label{stellar}
\mathcal{I}_{L}^{\star} = \mathcal{I}_L + \delta\mathcal{I}_L , \qquad \mathcal{S}_{L}^{\star} = \mathcal{S}_L + \delta\mathcal{S}_L ,
\end{equation}
with the tidal deformabilities defined in Eqs.~\eqref{tiddef}, \eqref{magdef}, \eqref{rottiddef} now directly related to quantities appearing in the metric by the associations

\begin{subequations}
\begin{align}
\lambda_2 &= 2 k_2^{\text{el}} R^5/3G , \\
\hat{\lambda}_2 &= 16 \mathfrak{k}^{\text{q}} m R^5/35c^2 , \\
\sigma_2 &= -k_2^{\text{mag}} m R^4/c^2 , \\
\hat{\sigma}_2 &= \mathfrak{f}^{\text{q}} m R^5/4c^2, \\
\lambda_3 &= 2 k_3^{\text{el}} R^7/15G , \\
\hat{\lambda}_3 &= 4 \mathfrak{k}^{\text{o}}  R^5/5 c^2 , \\
\sigma_3 &= -32k_3^{\text{mag}} m R^6/135 c^2 , \\
\hat{\sigma}_3 &= 8 \mathfrak{f}^{\text{o}} m R^5/15 c .
\end{align}
\end{subequations}
imposed by the equality of Eqs.~\eqref{diffmetric} and \eqref{Thornemetric}.\footnote{The signs in Eqs.~\eqref{tiddef}, \eqref{magdef}, \eqref{rottiddef} have been chosen so that the tidal deformabilities are positive. The linear tidal deformabilities have units of $[\text{length}]^{2\ell+1}/G$, and the rotational-tidal deformabilities have units of $[\text{length}]^{2\ell \pm 1}/G$, where the upper sign is for quadrupole moments and the lower is for octupole moments.} These expressions for the multipole moments should be equivalent to those calculated by Ref.~\cite{Pani_ext} using Ryan's method \cite{Ryan1,Ryan2} after accounting for differences in the definitions of the tidal moments and the Love numbers.

With the tidally induced multipole moments in hand, we write down the total multipole moments for a circular neutron-star binary to leading PN order, including the quadrupole correction identified by Ref.~\cite{Yagi_err}. Summing the orbital point-particle $\ell$-pole from Ref.~\cite{Wiseman} and the stellar $\ell$-poles, the total mass and current quadrupole and octupole moments for the system are

\begin{subequations} \label{totmulti}
\begin{align} 
\bar{\mathcal{I}}_{ab} &= \mu r^2 n_{\langle a} n_{b \rangle} + \frac{8}{9} \frac{m_2}{M} \left[2 v^c \epsilon_{cd \langle a} \mathcal{S}_{b \rangle d}^{\star} - r n^c \epsilon_{cd \langle a} \frac{d\mathcal{S}_{b \rangle d}^{\star}}{dt} \right] \nonumber \\ &\phantom{=}+ \mathcal{I}_{ab}^{\star} + (1 \leftrightarrow 2) ,  \\
\bar{\mathcal{S}}_{ab} &= -\mu \frac{(1-q)}{(1+q)}r^2 \epsilon_{cd\langle a} n_{b \rangle} n^c v^d + \mathcal{S}_{ab}^{\star} + (1 \leftrightarrow 2) , \\
\bar{\mathcal{I}}_{abc} &= -\mu\frac{(1-q)}{(1+q)} r^3 n_{\langle abc \rangle} + \mathcal{I}_{abc}^{\star} + (1 \leftrightarrow 2) , \\
\bar{\mathcal{S}}_{abc} &= \mu\left(1-\frac{3\mu}{M}\right)r^3 \epsilon_{ed\langle a} n_{bc \rangle} n^e v^d + \mathcal{S}_{abc}^{\star} + (1 \leftrightarrow 2) ,
\end{align}
\end{subequations}
where $(1 \leftrightarrow 2)$ indicates that the stellar multipoles from neutron star 2 can be obtained by reversing the labels on the masses, spins and tidal deformabilities.

%%%%%%%%%%%%%%%%%%%%%%%%%%%%%%%%%%%%%%%%%%
\section{Tidal phase}\label{sec:tidphase}
%%%%%%%%%%%%%%%%%%%%%%%%%%%%%%%%%%%%%%%%%%

Each of the stellar multipole moments calculated above contributes to the phasing of the binary waveform. The effect of the tidal mass multipole moments $-\lambda_{\ell} \mathcal{E}_L$ was computed by Refs.~\cite{Flanagan,Hinderer,Yagi}. The effect of the tidal current quadrupole moment $-\sigma_2 \mathcal{B}_{ab}$ was computed in Ref.~\cite{Yagi}. In this section, we revisit these calculations and add in the rotational-tidal deformations, as well as the gravitomagnetic octupole moment. We use geometrized $G=c=1$ units throughout the section, and label the masses, spins and tidal deformabilities according to the individual neutron star to which they refer (e.g.~$m_1$, $m_2$, $\chi^{(1)}$, $\chi^{(2)}$, $\lambda^{(1)}$, $\lambda^{(2)}$). We recall that $M$ is the total mass of the system and $\mu$ is the reduced mass.

The time-domain gravitational-wave phase $\Phi$ is related to the orbital frequency $\Omega$ via the formula  $d\Phi/dt = 2 \Omega$ (i.e.~the gravitational-wave frequency is twice the orbital frequency). Using the chain rule, one can re-express this relation as

\begin{equation}
\frac{d\Phi}{d\Omega} = 2\Omega \frac{dE/d\Omega}{dE/dt} ,
\end{equation}
where $E$ is the orbital energy and $dE/dt$ is the gravitational-wave energy flux. The phase $\Psi$ of the Fourier transform of the time-domain waveform is then given by \cite{Tichy}

\begin{equation} \label{stationaryphase}
\Psi = 2\Omega t_0 - 2\Phi_0 - \frac{\pi}{4} - 2 \int_{\Omega_0}^{\Omega} d\omega \, (\Omega-\omega) \frac{dE/d\omega}{dE/dt}
\end{equation}
in the stationary-phase approximation. Since the reference time $t_0$, phase $\Phi_0$ and angular frequency $\Omega_0$ are unobservable, the physically relevant part of Eq.~\eqref{stationaryphase} excludes the unknown linear function of $\Omega$ \cite{Tichy}, and can be obtained by integrating

\begin{equation} \label{GWphase}
\frac{d^2\Psi}{d\Omega^2} = 2\frac{d E/d\Omega}{dE/dt} 
\end{equation}
twice with respect to $\Omega$ \cite{Flanagan}. Thus, the phasing formula \eqref{GWphase} permits the tidal phasing to be computed, provided the tidal corrections to the $E$ and $dE/dt$ are known.

In the following subsections, we compute the tidal phase contributed by neutron star 1 only -- the expressions for $E$, $dE/dt$ and $\Psi$ omit terms generated by the deformation of neutron star 2. Such terms can, however, be obtained \textit{a posteriori} by swapping the labels 1 and 2 on the masses, spins and tidal deformabilities. The contributions from each star are then summed to give the total energy, gravitational-wave energy flux, and tidal phase.

\subsection{Conservative correction} \label{subsec:conservative}

The orbital energy is calculated from an integral of the binary's equation of motion. Corrections to the system's relative acceleration due to stellar multipoles were investigated by Ref.~\cite{Racine}, and a general 1PN equation of motion for a member of an $N$-body system is given in Eq.~(6.11) of that reference. Specializing to a binary, linearizing in spin and tidal deformations, and truncating at $\ell = 3$, the relative acceleration due to the multipoles raised on neutron star 1 reads

\begin{widetext}
\begin{align} \label{accel}
a_a = -\frac{M}{r^2} \bigg\{ n_a + \sum_{\ell=2}^3 (-1)^{\ell}\frac{(2\ell+1)!!}{\ell!} \frac{\mathcal{I}_L^{\star (1)}}{m_1} \frac{n_{\langle aL \rangle}}{r^{\ell}} &+ \sum_{\ell=1}^2 (-1)^{\ell} \frac{4(2\ell+1)!!}{\ell!(\ell+2)} \frac{1}{r^{\ell+1}} \bigg[ \epsilon_{cae}\delta^{eb} \frac{d\mathcal{S}^{cL}_{\star (1)}/dt}{m_1} r n_{\langle bL \rangle} \nonumber \\ &-(2\ell+3) v^d \frac{\mathcal{S}^{cL}_{\star (1)}}{m_1} \left( \epsilon_{cae} \delta^{eb} n_{\langle bdL \rangle} + \epsilon_{dce} \delta^{eb} n_{\langle abL \rangle} \right) \bigg] \bigg\} .
\end{align}
\end{widetext}
Here, we have suppressed terms proportional to the spins, since they do not have a tidal origin and they cannot combine with the rotational-tidal terms at leading order. We have retained terms proportional to mass multipoles up to 0PN, and terms proportional to current multipoles up to 1PN, since we are interested in only the leading-order tidal terms. The acceleration due to the multipoles of neutron star 2 can be obtained by exchanging the labels 1 and 2.

Inserting the stellar multipoles of Eq.~\eqref{stellar}, we obtain for the radial component $a_r = n^a a_a$ of the acceleration

\begin{align} \label{arad}
a_r = -\frac{M}{r^2} \bigg[ 1 &+ \frac{3 m_2 (3\lambda^{(1)} + 8 v^2 \sigma^{(1)})}{m_1 r^5} \nonumber \\ &- \chi^{(1)}\frac{6 m_2 v (18\hat{\lambda}_2^{(1)} - 2\hat{\lambda}_3^{(1)} + 4 \hat{\sigma}_2^{(1)} - 3 \hat{\sigma}_3^{(1)})}{m_1 r^6} \nonumber \\ &+ \frac{60 m_2 (\lambda_3^{(1)} + 9 v^2 \sigma_3^{(1)})}{m_1 r^7} \bigg]
\end{align}
in an expansion in powers of $1/r$. We observe that the tidal quadrupole deformabilities enter at the lowest order, and that the gravitomagnetic deformabilities are suppressed by a factor of $v^2$ (1PN) relative to their gravitoelectric counterparts. The rotational-tidal deformabilities make their appearance at an intermediate order between the tidal quadrupoles and octupoles. This hierarchy will manifest itself repeatedly in the results of this section, and it is reflected in the ultimate PN scaling of the phase, Eq.~\eqref{psifinal}.

Given the radial acceleration, the equation of motion $a_r =- r \Omega^2$ can be solved order by order for the radius $r$ of the circular orbit. Exchanging $v$ for the PN parameter $x= v^2$, we find

\begin{align} \label{rsol}
r = \frac{M^{1/3}}{\Omega^{2/3}} \bigg[ 1 &+ \frac{3 q \lambda^{(1)}}{M^5}x^5 + \frac{8 q \sigma^{(1)}}{M^5} x^6  \nonumber \\ &- \chi q \frac{36\hat{\lambda}_2^{(1)} -4 \hat{\lambda}_3^{(1)} + 8\hat{\sigma}_2^{(1)} -6 \hat{\sigma}_3^{(1)}}{M^6}x^{6.5} \nonumber \\ &+ \frac{20q \lambda_3^{(1)}}{M^7} x^7 + \frac{180 q \sigma_3^{(1)}}{M^7} x^8 \bigg] .
\end{align}

Integrating the equation of motion with respect to $r$, and plugging in the solution Eq.~\eqref{rsol}, we obtain the following expression for the corrected orbital energy:

\begin{align} \label{energy}
E = -\frac{1}{2} \mu x \bigg[ 1 &- \frac{9q\lambda^{(1)}}{M^5} x^5 - \frac{24 q \sigma^{(1)}}{M^5}x^6 \nonumber \\ &+ \chi^{(1)}\frac{44q (18\hat{\lambda}_2^{(1)} -2 \hat{\lambda}_3^{(1)} + 4\hat{\sigma}_2^{(1)} -3 \hat{\sigma}_3^{(1)})}{7 M^6} x^{6.5} \nonumber \\ &- \frac{65 q \lambda_3^{(1)}}{M^7} x^7 - \frac{585q \sigma_3^{(1)}}{M^7} x^8 \bigg] .
\end{align}
This expressions leaves out next-to-leading order contributions from each of the tidal moments, as well as known PN point-particle terms. One observes that the gravitoelectric quadrupole correction enters at 5PN; the gravitomagnetic quadrupole follows at 6PN; and then all four rotational-tidal multipoles enter at 6.5PN. The octupole corrections come in at 7PN and 8PN for the gravitoelectric and gravitomagnetic deformabilities, respectively.

\subsection{Dissipative correction} \label{subsec:dissipative}

The general expression for the gravitational-wave power radiated due to mass and current $\ell$-poles is \cite{Thorne}

\begin{align} \label{GWpower}
\frac{dE}{dt} =& - \sum_{\ell=2}^{\infty} \frac{(\ell+1)(\ell+2)}{(\ell-1)\ell} \frac{1}{\ell!(2\ell+1)!!} \left\langle \frac{d^{\ell+1} \bar{\mathcal{I}}_L}{dt^{\ell+1}} \frac{d^{\ell+1} \bar{\mathcal{I}}^L}{dt^{\ell+1}} \right\rangle \nonumber \\
&- \sum_{\ell=2}^{\infty} \frac{4\ell(\ell+2)}{(\ell-1)} \frac{1}{(\ell+1)!(2\ell+1)!!} \left\langle \frac{d^{\ell+1} \bar{\mathcal{S}}_L}{dt^{\ell+1}} \frac{d^{\ell+1} \bar{\mathcal{S}}^L}{dt^{\ell+1}} \right\rangle ,
\end{align}
where angular brackets denote orbit averaging and the overall minus sign indicates an outgoing energy flux. Because the stellar multipoles are small relative to the orbital ones, it is the cross terms in Eq.~\eqref{GWpower} that are responsible for the bulk of the correction to the radiated power. For our purposes, the sums are truncated at $\ell=3$.

We calculate the modification of the binary's gravitational-wave energy flux by inserting the system's total multipoles \eqref{totmulti} (excluding, for now, the contribution from neutron star 2). We find

\begin{widetext}
\begin{align} \label{power}
\frac{dE}{dt} = -\frac{32 \mu^2}{5M^2}x^5 \bigg[ 1 &+ \frac{6(1+3q)\lambda^{(1)}}{M^5}x^5 - \frac{(1-113q)\sigma^{(1)}}{3M^5}x^6 - \chi^{(1)}\frac{36(5+17q)\hat{\lambda}_2^{(1)} - (1-113 q) \hat{\sigma}_2^{(1)} - 48q \hat{\lambda}_3^{(1)} - 72q \hat{\sigma}_3^{(1)} }{3M^6}x^{6.5} \nonumber \\ &+ \frac{80q\lambda_3^{(1)}}{M^7}x^7 + \frac{720q \sigma_3^{(1)}}{M^7} x^8 \bigg] .
\end{align}
\end{widetext}
One can see that the PN scaling of the various contributions is the same as in Eq.~\eqref{energy}. As above, only the 0PN point-particle term and the leading-order terms proportional to each tidal deformability have been retained.

\subsection{Phase} \label{subsec:phase}

Combining the results of Eqs.~\eqref{energy} and \eqref{power}, we compute the tidal phase correction via Eq.~\eqref{GWphase}. Because the PN scalings of the energy and gravitational-wave power corrections are the same, both effects contribute to the phasing of the waveform. Expanded in powers of the PN parameter $x$, the expression for the phasing due to a single neutron star is

\begin{widetext} 
\begin{align}\label{psi1}
\Psi = \frac{3M}{128 \mu}x^{-2.5} \bigg[ 1 &- \frac{24(1+12q)\lambda^{(1)}}{M^5} x^5 + \frac{10(1-617q)\sigma^{(1)}}{21 M^5}x^6 \nonumber \\ &+ \chi^{(1)} \frac{36(35+614q)\hat{\lambda}_2^{(1)} - (7-4751q)\hat{\sigma}_2^{(1)} - 2316q \hat{\lambda}_3^{(1)} - 3474q \hat{\sigma}_3^{(1)}}{21M^6}x^{6.5} \nonumber \\ &- \frac{4000q \lambda_3^{(1)}}{9 M^7} x^7  - \frac{29925 q \sigma_3^{(1)}}{11 M^7}x^8 + (1 \leftrightarrow 2)  \bigg] .
\end{align}
\end{widetext}
The contribution from the companion can be obtained by reversing the labels 1 and 2 on the masses and tidal deformabilities, as was done to produce the final expression given in Eq.~\eqref{psifinal}. We note that the PN scalings from above have carried through to the phase. As mentioned before, this expression neglects PN corrections apart from the leading-order terms involving each tidal deformability.

While the $\lambda$ and $\lambda_3$ phase terms in Eq.~\eqref{psi1} agree with the literature \cite{Yagi}, we note that the coefficient of $q$ in the $\sigma$ term is in disagreement with Ref.~\cite{Yagi_err}. This discrepancy is discussed in further detail below.

%%%%%%%%%%%%%%%%%%%%%%%%%%%%%%%%%%%%%%%%%%
\section{Discussion}\label{sec:disc}
%%%%%%%%%%%%%%%%%%%%%%%%%%%%%%%%%%%%%%%%%%

In this work, the $\ell \leq 3$ multipole moments of a slowly rotating neutron star deformed by a quasi-stationary tidal field were calculated using a simple prescription based on the multipole structure of the generic stationary, asymptotically flat vacuum metric of Ref.~\cite{Thorne}. In particular, the amplitudes of four separate rotational-tidal deformations generated by bilinear couplings between neutron star spin and the moments of the external tidal field were related to the Love numbers calculated in Refs.~\cite{Pani,Landry_int,GagnonBischoff}. The impact of these rotational-tidal deformations on the waveform phase $\Psi$ was calculated, with the result displayed in Eq.~\eqref{psifinal}. The spin-tide couplings were found to contribute a new 6.5PN term proportional to the composite tidal parameter $\tilde{X}$ defined in Eq.~\eqref{X}. In addition to recovering the known 5PN and 7PN phasing terms proportional to $\tilde{\Lambda}$ and $\tilde{\Lambda}_3$, the 8PN gravitomagnetic octupole term proportional to the effective deformability $\tilde{\Sigma}_3$ defined in Eq.~\eqref{sigma3} was calculated for the first time. The rotational-tidal phase is thus larger than the phasing due to octupole tides, and it will be enhanced in systems with significant spin, since it scales linearly with the dimensionless spins of the individual neutron stars.

The discrepancy in the $\tilde{\Sigma}$ phase term reported above is unexpected and concerning. As this work was nearing completion, it was pointed out \cite{comm} to the author that it results from the procedure used to calculate the energy: an integration of the equation of motion can't account for non-conservative contributions associated with the gravitomagnetic field. Accordingly, a discrepancy of similar origin may affect the phase coefficients for $\sigma_3$, $\hat{\sigma}_2$ and $\hat{\sigma}_3$. Nonetheless, although the numerical values of the coefficients may change, the PN scaling and overall form of these terms is expected to be robust. The issue will be investigated in further detail.

A number of studies have investigated the systematic errors introduced into Bayesian parameter estimation when waveform models omit certain PN terms \cite{Favata, Wade, HardLove}, such as the unknown 4PN point-particle phasing and next-to-leading order tidal corrections. It may be worthwhile to perform a similar analysis to determine whether omission of the effective rotational-tidal parameter $\tilde{X}$ can bias recovery of the 5PN tidal parameter $\tilde{\Lambda}$ by the Advanced LIGO \cite{LIGO} and VIRGO \cite{VIRGO} interferometers or, more likely, the 6PN tidal parameters $\delta \tilde{\Lambda}$ or $\tilde{\Sigma}$ by future-generation detectors such as LISA \cite{LISA}, Einstein Telescope \cite{ET} or Cosmic Explorer \cite{CE}. One could also evaluate the prospect of measuring $\tilde{X}$ itself with these planned detectors. The natural starting point for this investigation would be a Fisher matrix study \cite{Cutler}.

Finally, we remark that this calculation merely sought the leading-order rotational-tidal phase terms, and did not attempt a consistent derivation of all tidal terms up to 6.5PN (let alone 8PN, the highest order featuring in Eq.~\eqref{psifinal}). While it may be desirable to understand the next-to-leading- and higher-order corrections to the various tidal phase contributions, the development of a fully consistent phase formula up to 6.5PN is principally limited by our ignorance of point-particle terms starting at 4PN. Since these point particle terms enter at relatively low PN order, it is likely that the uncertainty they introduce in the recovery of higher-order parameters like $\tilde{\Sigma}$ or $\tilde{X}$ outweighs the corrections from next-to-leading-order tidal terms.

%%%%%%%%%%%%%%%%%%%%%%%%%%%%%%%%%%%%%%%%%%
\acknowledgments
%%%%%%%%%%%%%%%%%%%%%%%%%%%%%%%%%%%%%%%%%%

The author thanks Bob Wald, Reed Essick and Eric Poisson for useful comments about the manuscript. This work was supported in part by the Natural Sciences and Engineering Research Council of Canada, and by NSF grants PHY 15-05124 and PHY 17-08081 to the University of Chicago.

%%%%%%%%%%%%%%%%%%%%%%%%%%%%%%%%%%%%%%%%%%
\bibliography{multipoles-refs}
%%%%%%%%%%%%%%%%%%%%%%%%%%%%%%%%%%%%%%%%%%

\end{document}